\documentclass[aps,prb,showpacs,twocolumn,floatfix]{revtex4}
\usepackage{graphics}

\begin{document}

\title{Cross-sectional STM of Mn-doped GaAs: theory and experiment}
\author{J. M. Sullivan, G. I. Boishin, L. J. Whitman, A. T. Hanbicki, 
B. T. Jonker, and S. C. Erwin}
\affiliation{Naval Research Laboratory, Washington, D.C. 20375}
\date{\today}

\begin{abstract}
We report first-principles calculations of the energetics and
simulated scanning tunneling microscopy (STM) images for Mn dopants 
near the GaAs (110) surface, and compare the results with cross-sectional STM
images.  The Mn configurations considered here include
substitutionals, interstitials, and complexes of substitutionals and
interstitials in the first three layers near the surface.  Based on
detailed comparisons of the simulated and experimental images, we
identify three types of Mn configurations imaged at the surface: (1)
single Mn substitutionals, (2) pairs of Mn substitutionals, and (3) 
complexes of Mn substitutionals and interstitials.
\end{abstract}
\maketitle

\section{Introduction}\label{Introduction}
GaAs can be doped with Mn to form a dilute magnetic semiconductor
with a Curie temperature as high as 160 K. \cite{chiba03,ku03,edmonds02}
It is generally accepted that holes created by the 
substitution of Mn for Ga mediate the ferromagnetic
interaction between Mn dopants in this material.\cite{dietl00,dietl01} 
Naively, each substitutional Mn is 
expected to produce one hole; therefore the nominal hole concentration, {\it p}, 
should be equal to the Mn concentration.  For typical Mn fractions of
5\%\/, this
implies a hole concentration $p = 1.1\times10^{21}$ cm$^{-3}$.  Measured hole
concentrations of as-grown material are much less than this, typically
by a factor $\sim$3.  The source of this compensation remains somewhat
controversial, having been variously attributed to either excess As in the
form of antisites and interstitials\cite{sanvito02,bergqvist03,hayashi01,potashnik01} 
or Mn interstitials. \cite{masek01,yu02,erwin02,mahieu03}
Recent experiments show that careful annealing near the growth temperature
can significantly enhance the conductivity and hole concentration.\cite{yu02,edmonds02} 
This, as well as recent ion channelling experiments,\cite{yu02}
suggests that interstitial Mn is the more 
likely source. A complete picture is lacking, however; for example, the 
distribution of Mn interstitials in as-grown material is still unclear.

Cross-sectional scanning tunneling microscopy (XSTM) is an effective
tool for addressing this issue because it can image, with atomic resolution, 
the structural and electronic configuration of impurities and defects 
as they are present in the bulk.
There are several recent XSTM studies of 
Mn-doped GaAs;\cite{grandidier00,mahieu03,tsuruoka01,tsuruoka02} these 
studies used the correlation between transport data, level of Mn 
doping, and observed defect density to infer the location 
(substitutional versus interstitial) of Mn dopants and As antisites. 
The relative abundance of As antisites and interstitial Mn deduced by
these studies lacks a clear consensus: one study finds that
antisites are not present at all,\cite{tsuruoka01} while another finds
that both antisites and interstitial Mn are present,
and compensate substitutional Mn.\cite{mahieu03}

To investigate the nature of dopants and defects in this material, we
performed high resolution XSTM measurements on a (110) cleavage plane
of Mn-doped GaAs and used first-principles calculations to interpret
the images. Specifically, we used density-functional theory to simulate STM
images for a number of Mn configurations near
the GaAs (110) surface.  Configurations were chosen based on their
calculated energetic stability, and included several metastable
configurations that would probably be kinetically stabilized. A detailed
comparison of the resulting simulated and experimental images
leads us to the following conclusions: (1) isolated substitutional 
Mn as well as pairs of substitutional Mn occur with roughly comparable
frequency in bulk Mn-doped GaAs, and (2) interstitial Mn is typically 
bound in complexes with substitutional Mn in several different configurations.

\section{Experimental XSTM Images}
The Mn-doped GaAs sample used in this study was grown by molecular-beam
epitaxy (MBE) using well established methods and conditions.\cite{ohno96}  
A buffer layer of high quality undoped GaAs was first grown on an {\it n}-type 
GaAs substrate. This was followed by growth of a 100 nm thick GaAs buffer
layer grown at 250$^{\circ}$ C and a 260 nm thick layer of Mn-doped GaAs at 
the same temperature. Growth quality was monitored by reflection high-energy 
electron diffraction.  The Mn concentration was determined with x-ray diffraction 
to be 0.6\%\/.

This sample was then characterized by XSTM
measurements in a multichamber 
ultrahigh-vacuum facility. It order to obtain an atomically abrupt 
cleave across the epilayer, the samples were thinned 
{\it ex situ} to $\leq$200 $\mu$m. After being loaded into the XSTM chamber 
(base pressure $\leq$10$^{-10}$ Torr), the samples were scribed and cleaved {\it in situ} 
to expose a (110) surface (perpendicular to the MBE growth direction), as 
previously described.\cite{nosho00}  The constant-current (40 pA) images 
shown here are of filled electronic states measured with a sample bias of 
$-$2.5 V.

\begin{figure}
\resizebox{8cm}{!}{\includegraphics{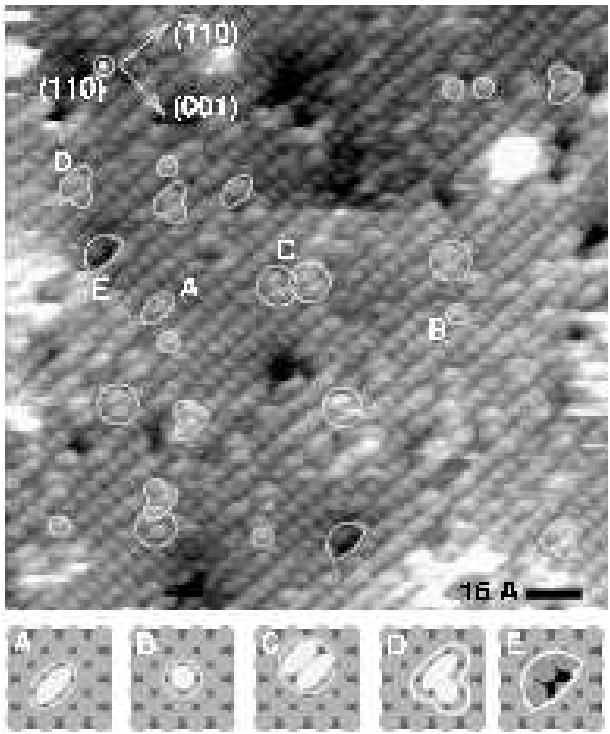}}%
\caption{\label{expt_stm} 
Constant-current filled-state XSTM image of a (110) cleavage plane of Mn-doped
GaAs (001). Five reproducible features (A-E) are marked on the image and drawn
schematically below.
} 
\end{figure}

Figure~\ref{expt_stm} shows the results of filled-state XSTM measurements on
this sample.
The scan area of this image is 170 $\times$ 170 \AA$^{2}$ and displays $\sim$1500 
surface atoms. Filled state STM images of GaAs (110) reveal the As surface sublattice;
the surface Ga sublattice evident in empty-state images, on the other hand, 
is not directly revealed in this image. Since only the top As atoms contribute, 
STM of the GaAs (110) surface reveals only every other layer in the (001) direction.

Figure~\ref{expt_stm} reveals many surface features not normally seen in STM
of the GaAs (110) surface. 
Several occur with enough regularity to warrant close attention. Feature A is an
elliptical region of intensity in an As (001) plane with major and
minor axes the size of two and one As surface atoms, respectively.
Feature B is circular, and only slightly larger and more intense 
than the contributions from the surrounding As. 
Feature C is more diffuse and broad than both A and B, with
intensity spread over two pairs of As atoms in neighboring (001) planes.
As discussed in more detail below, this is 
what one would expect if there were two A features in neighboring
As (001) planes. Feature D is of similar shape to that of C but is
more asymmetric, with one side of the defect the size of a single surface As 
atom. There does not appear to be a preferred crystalline orientation of
feature D as different orientations are 
visible in Fig.~\ref{expt_stm}. Feature E is a dark region close to a
surface As atom displaced in the (00$\overline{1}$) direction;
there is also an overall apparent depression in the vicinity of this
feature, characteristic of band bending that occurs near positively charged 
defects on a {\it p}-type surface.\cite{ebert99}

On the basis of the experimental data alone it is extremely difficult, 
if not impossible, to determine what dopants or defects produce 
these features. As we show in this work, however, when interpreted with
first-principles theory, these measurements resolve both the type
and crystalline orientation of Mn dopants in GaAs.

\section{Theoretical Approach}
\subsection{Physical and Electronic Structure}
We modeled the GaAs (110) surface in a supercell slab geometry consisting 
of 5 layers of GaAs in a 4$\times$4 surface unit cell. A vacuum 
region of 13.4 \AA\ was used for all the calculations. In such a cell
the separation between a dopant and its periodic image is $\sim$16
\AA\/, sufficiently large to ensure that dopants interact only negligibly 
with their periodic images. The bottom layer of the slab was passivated; 
the atomic positions were fully relaxed in all but the bottom two layers.

The wavefunctions and charge density were expanded in a plane-wave basis
and evaluated using an ultrasoft pseudopotential approach\cite{vanderbilt90,kresse94} 
as implemented in the VASP code.\cite{kresse96a,kresse93,kresse96}
Electron correlations were treated at the level of the local spin-density
approximation (LSDA) with total energies converged to 10$^{-4}$ eV.

A single wave vector was used to sample the Brilloun zone of the slab supercells
and a plane-wave cutoff of 227.24 eV was used for all the calculations.
For supercells containing Mn, atoms within $\sim$7.0 \AA\ of the Mn 
site(s) were fully relaxed until the total energy between two structural configurations
changed by less than 10$^{-3}$ eV.

\subsection{Simulated STM Images}
Theoretical STM images were simulated using the method of Tersoff and Hamann.\cite{tersoff83} 
In this approach, the central quantity is the the local density of states (LDOS) 
near the surface. The LDOS is integrated over an energy range determined by the 
experimental bias voltage; contours of constant energy-integrated LDOS simulate
a constant-current image. For filled state 
imaging, the upper bound of this energy range is the Fermi level. The Mn-doped GaAs 
sample under consideration in this work is {\it p}-type with an estimated hole density 
of 10$^{19}$cm$^{-3}$, so that the Fermi level 
should be near the valence band maximum (VBM). Since we do not 
know the distribution of donors and acceptors present in the sample, this information 
alone is not sufficient to determine the exact location of the Fermi level for our
sample. However, we have found that, as a consequence of the large bias,
the simulated images are rather insensitive to the value of the Fermi level used 
in the LDOS integration, and so
we choose arbitrarily a point 0.4 eV above the VBM as the upper bound for all
the STM simulations. For this choice, the
Fermi level is above the acceptor level of substitutional Mn and below the 
donor levels of interstitial Mn. Simulations performed with the position of the Fermi 
level shifted by $\pm$0.3 eV give similar results. The location of the VBM for each 
type of dopant was determined by inspection of the total DOS relative to that 
projected onto the dopant site.

As a measure of the agreement between simulated and measured STM images, we consider mainly
the overall shape and spatial extent of features of interest. The defect-free (110) 
surface has a ($\overline{1}$10) mirror plane symmetry, so it is also useful to note 
whether a particular dopant configuration preserves, even approximately, this symmetry. Finally, we
use qualitative arguments to judge whether the number of observed
features are consistent with our total-energy calculations.

\section{Substitutional Mn}
\subsection{Simple Substitutional Dopants}
\begin{figure*}
\resizebox{16cm}{!}{\includegraphics{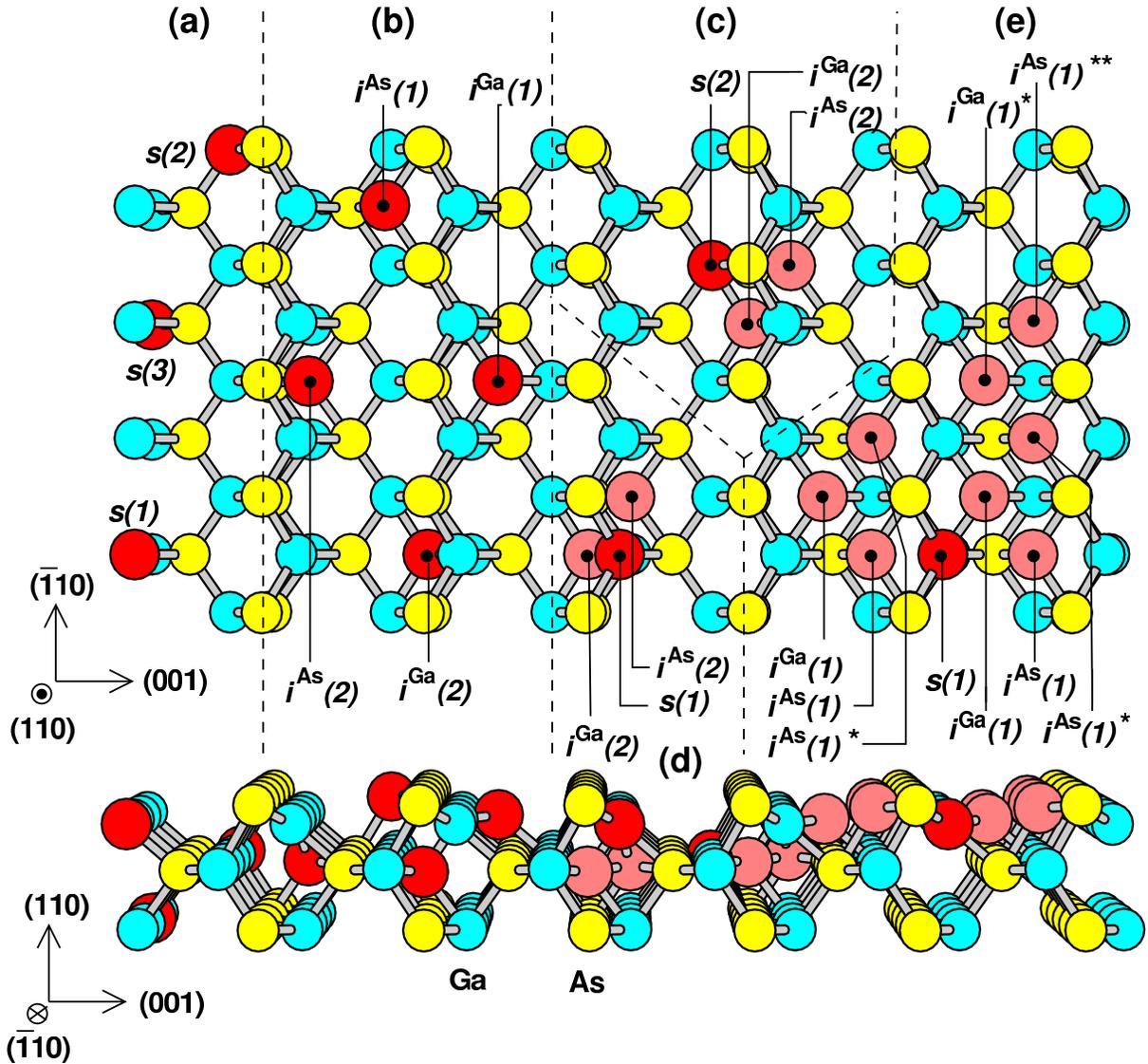}}%
\caption{\label{struk_new} (color) Top view (upper) and side view (lower) showing the
Mn dopant configurations considered in this work. Ga, As, and Mn atoms are shown as
cyan, yellow, and red circles, respectively. The dashed lines separate regions containing
different types of dopants. From left to right: (a) substitutional
Mn in the first three layers, (b) interstitial Mn in the first 
two layers, (c) complexes of substitutional and interstitial Mn in layer 2, 
(d) complexes of substitutional Mn in layer 1 and interstitial Mn 
in layer 2, (e) complexes of substitutional and interstitial
Mn both in layer 1. For the complexes in regions (c), (d), and (e), several alternative
locations of the interstitial Mn involved in the complex are shown in a lighter shade 
of red. See text for explanation of the notation.} 
\end{figure*}

Substitutional Mn dopants are of primary importance for the magnetic and 
transport properties of Mn-doped GaAs, and hence we discuss the simulated STM 
images for these types of dopants first. Since STM is a near-surface sensitive 
probe, we have only considered substitutional Mn dopants in the top three layers 
near the (110) surface, as shown in Fig.~\ref{struk_new}(a).

\begin{figure}
\resizebox{8cm}{!}{\includegraphics{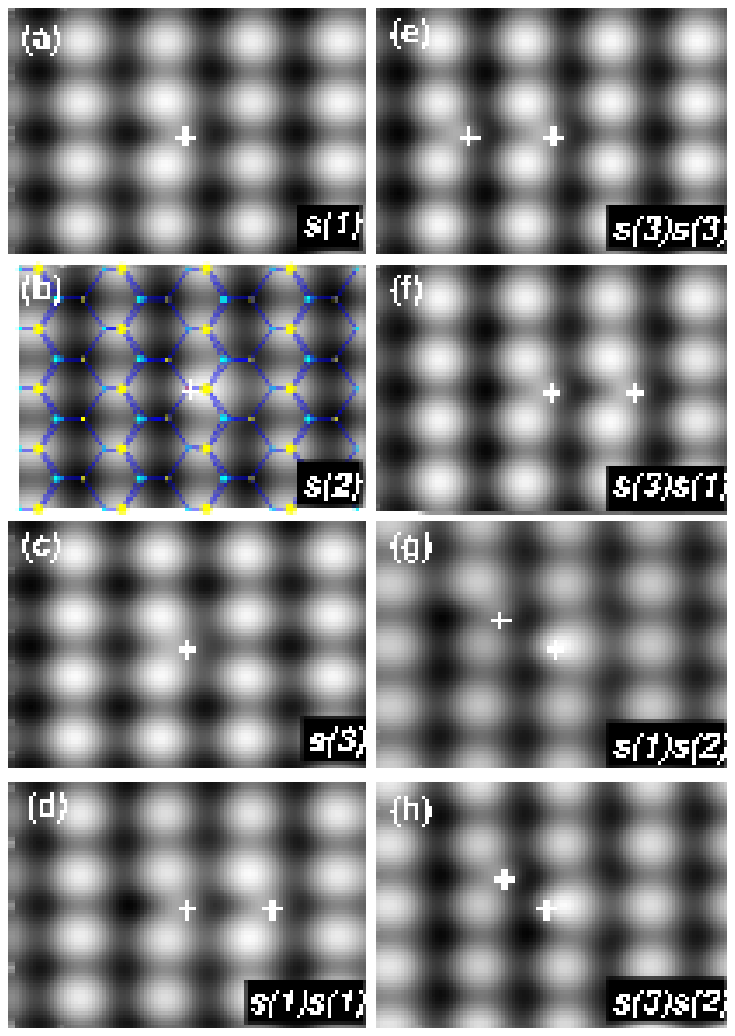}}%
\caption{\label{mnga_stms_combined} Panels (a), (b), and (c) show show simulated STM
images of isolated Mn substitutionals. 
Panels (d) through (h) show simulated images for Mn substitutional pairs. 
Mn positions are marked with {\bf +}. The crystalline orientation is the same 
as Fig.~\ref{struk_new}.
} 
\end{figure}

We will use the notation $s(n)$ to denote a substitutional Mn in layer $n$ (with
the topmost surface layer defined as $n=1$). Figures~\ref{mnga_stms_combined}(a),
(b), and (c) show the simulated images from $s(1)$, $s(2)$, and $s(3)$, respectively.
The $s(1)$ and $s(3)$ images are strikingly similar, for the following reasons.
The $s(1)$ Mn atom forms bonds with two As atoms that are nominally in 
the same plane; likewise for the $s(3)$ Mn atom. Thus the overall shape of the dopant 
consists 
of three overlapping maxima, centered on the two As atoms and on the Mn atom, 
giving rise to an elongated feature with a ($\overline{1}$10) mirror-plane symmetry. 
Hence, both of these sites are consistent with feature A in Fig.~\ref{expt_stm}.

The $s(2)$ dopant on the other hand, has a very different image. 
In the $s(2)$ location the Mn forms a bond
with an As atom in the top layer, with the bond pointing toward the surface.
The perturbation due to $s(2)$ is centered on the surface As site involved in the bond. 
This overlaps with the contribution from the Mn dopant itself, and thus the 
shape is more circular compared to the $s(1)$ and $s(3)$ dopants. It is also
larger in size than surface As atom contributions and has a 
($\overline{1}$10) mirror-plane symmetry. These agree qualitatively
with the characteristics of feature B in Fig.~\ref{expt_stm}, leading
us to identify those features as arising from $s(2)$ dopants.

\subsection{Substitutional Mn Pairs}
Since there will always be some fraction of neighboring Mn that occur
simply by chance, we consider pairs of substitutional Mn 
as well. Figures~\ref{mnga_stms_combined}(d), (e) and (f) show simulated images of 
pairs of Mn substitutionals $s(1)s(1)$, $s(3)s(3)$ and $s(3)s(1)$, respectively.
For the $s(3)s(1)$ pair, we note that even though the two Mn involved are
$\sim$6.8 \AA\ apart, and in different atomic planes, because of the projected 
view of the STM scan the appearance of this pair is
quite similar to those of the $s(1)s(1)$ and $s(3)s(3)$ pairs. 
All three of these simulated images have a shape 
comparable to that of feature C in Fig.~\ref{expt_stm}. Common features
include a region of low intensity between the two dopants, a ($\overline{1}$10) 
mirror-plane symmetry (the experimental data appear to only approximately have 
this symmetry) and a similar spatial extent involving four surface As atoms. 
It is also interesting to note that these pairs have simulated
images that are essentially a linear superposition of the images of the isolated
substitutionals shown in Figs.~\ref{mnga_stms_combined}(a) and (c). This suggests that,
at the level of detail that STM reveals, two substitutional Mn 
do not interact very strongly when they are next-nearest neighbors.

Figures~\ref{mnga_stms_combined}(g) and (h) show the simulated images for
nearest-neighbor substitutional pairs $s(1)s(2)$ and $s(3)s(2)$, respectively. 
Overall, the simulated images of these pairs are similar to what one would
obtain from a linear superposition of the images of the isolated defects.
However, the simulated image for $s(1)s(2)$ clearly reveals that the contribution 
from the $s(1)$ dopant is now asymmetric, lacking a ($\overline{1}$10) mirror plane. 
This indicates that substitutional pairs interact more strongly when they are
in near-neighbor configuration than in the next-nearest-neighbor configuration
described in the paragraph above.
The shape of these simulated images is generally consistent with
that of feature D in Fig.~\ref{expt_stm}, with a characteristic ``L'' shape
involving three As atoms, two of which are in the same As (001) plane. We
have not found any other dopants that yield such a favorable comparison to
the experimental data, leading us to identify feature D as $s(1)s(2)$ and
$s(3)s(2)$ pairs. In the experimental data one can observe three of the
four possible crystal orientations of these pairs of substitutionals.

An estimate of the number of neighboring pairs of Mn substitutionals based 
on a random distribution of Mn on the Ga sublattice is much lower than the 
number of compound features, in particular those labeled ``C'', seen in 
Fig.~\ref{expt_stm}. Even if there is some positional correlation 
among the substitutional dopants, for example, the   
strong short-range attraction between Mn found in a recent theoretical study,\cite{vanschilfgaarde01}
it is unlikely that a description based entirely
on substitutionals and clusters of substitutionals can describe
the frequency and shapes of all the features in 
Fig.~\ref{expt_stm}. This suggests that other Mn complexes, in particular those 
involving interstitial Mn, may help explain the remaining features.

\section{Dopant Complexes}
\subsection{Total Energy Considerations}
It has recently been shown that although the formation energy of interstitial 
Mn is much higher than that of substitutional Mn in bulk GaAs, the presence of 
the (001) growth surface significantly enhances the likelihood of Mn occupying 
interstitial sites.\cite{erwin02} Thus one expects that both
substitutional and interstitial Mn will be present in typical Mn-doped
GaAs samples, as shown by recent measurements of the Mn distribution
in Mn-doped GaAs.\cite{yu02}

In the bulk, the barrier for diffusion of a charged ($+$2{\it e}) interstitial 
Mn is $\sim$0.5 eV; 
therefore at room temperature we expect interstitial Mn to rapidly
diffuse throughout the bulk material unless it becomes kinetically 
trapped. Since substitutional dopants are acceptors,
they are a natural trap for interstitials, electrostatically binding 
to them in a number of different physical configurations.
Clustering phenomena for compensating dopants
such as these have recently been predicted using Monte Carlo techniques.\cite{timm02}
On the other hand, since interstitial Mn in bulk is a double donor with 
charge of $+$2{\it e}, under conditions of filled-state 
imaging (for which the tip is positively biased relative to the sample) 
free interstitial Mn will be repelled away from the surface, leaving behind 
only those interstitials that are bound in complexes. 
To determine which such interstitial-substitutional complexes are most 
likely to be present, we turn to total-energy calculations.

To simplify the discussion, we extend the notation introduced earlier
to now include Mn interstitials, and complexes of interstitial and substitutional Mn.
Isolated interstitial Mn in layer $n$, coordinated to As and Ga,
will be referred to as $i^{\rm As}(n)$ and $i^{\rm Ga}(n)$, respectively.
For reference, the configuration of isolated interstitial dopants in the first
two layers is shown in Fig.~\ref{struk_new}(b). Complexes composed of 
interstitial and substitutional Mn will be referred to with a 
composite notation that takes into account the crystalline orientation 
of the atoms involved. For example, since the GaAs (110) surface 
does not have a (001) mirror-plane symmetry, there are two distinct 
ways of forming a complex of interstitial and substitutional Mn. 
In Fig.~\ref{struk_new} the surface Ga atoms are to the 
right of the surface As atoms to which they are bonded, so that we may 
consider complexes where the interstitial Mn is to the left or right
of the neighboring substitutional dopant; such complexes will be denoted as 
$i^{\rm As}(n)s(m)$ and $s(m)i^{\rm As}(n)$, respectively. 
Figure~\ref{struk_new}(c), (d), 
and (e) shows the configuration of near-surface complexes 
we have considered in this work. Thus, for example, Fig.~\ref{struk_new}(c) 
shows the complexes $s(2)i^{\rm Ga}(2)$ and $s(2)i^{\rm As}(2)$, which
we discuss below.

\begin{figure}
\resizebox{8cm}{!}{\includegraphics{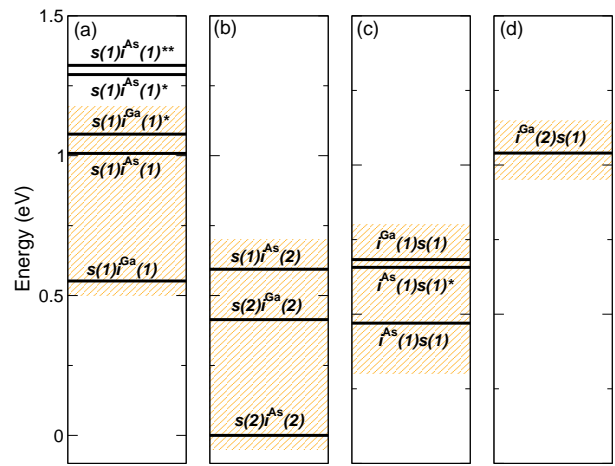}}%
\caption{\label{defect_energies} Relative total energies of complexes of 
interstitial and substitutional Mn near the GaAs (110) surface: (a) 
$s(1)i^{\rm Ga}(1)$ and $s(1)i^{\rm As}(1)$, (b) $s(m)i^{\rm Ga}(2)$
and $s(m)i^{\rm As}(2)$ ($m=1,2$), (c) $i^{\rm Ga}(1)s(1)$ and 
$i^{\rm As}(1)s(1)$, and (d) $i^{\rm Ga}(2)s(1)$. Simulated STM images 
are provided for the low-energy configurations in the shaded regions.
}
\end{figure}

Figure~\ref{defect_energies} shows the calculated total energies of Mn 
complexes relative to that of the $s(2)i^{\rm As}(2)$ dopant, which
we identify as the lowest energy configuration. There is a large 
variation in the energies, up to $\sim$1.3 eV above the energy of
$s(2)i^{\rm As}(2)$. We also note that for complexes involving
only Mn atoms on the surface (Figs.~\ref{defect_energies}(a) and (c)) 
the lowest energy configuration occurs when the interstitial site 
is nearest the substitutional site. Relative to these nearest-neighbor
complexes, when interstitial Mn is displaced by lateral translations
in the ($\overline{1}$10) direction, the energy increases dramatically;
the next lowest energy configurations are $\sim$0.3 eV higher in energy. 
Thus, diffusion of surface interstitial Mn from high-energy locations 
(denoted with $^{\rm *}$ and $^{\rm **}$) to low-energy configurations 
is likely to form nearest-neighbor complexes on the surface.

It is difficult to estimate, on energetic grounds alone, the expected 
number of such complexes. The presence of both the surface and the STM tip
create additional complications. For example,
the energy of interstitial Mn depends strongly on its 
proximity to the GaAs (110) surface; an interstitial Mn in the interior
of GaAs far from the surface can lower its energy as much as 
$\sim$1 eV by moving to a location just below the surface.\cite{sullivan03} 
This suggests that interstitial Mn will diffuse from the bulk-like region 
toward the surface; during the STM scans these ``excess'' interstitials 
will be observed only if they are bound in complexes. Such phenomena
are difficult to quantify and beyond the scope of this work. Instead we
choose simply to report simulated STM images of several low-energy 
complexes within the shaded regions of Fig.~\ref{defect_energies}.

\subsection{Simulated STM Images}
\subsubsection{Substitutional-interstitial complexes}
\begin{figure}
\resizebox{8cm}{!}{\includegraphics{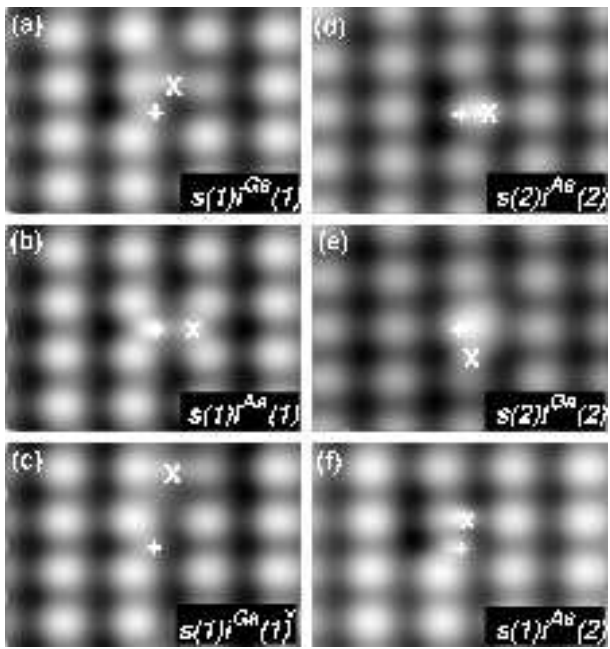}}%
\caption{\label{mnga_si_combined} Simulated STM images for Mn substitutional-
interstitial complexes. 
The locations of substitutional and interstitial
Mn are denote with $+$ and $\times$, respectively. The crystalline 
orientation is the same as in Fig.~\ref{struk_new}.
} 
\end{figure}

Figure~\ref{mnga_si_combined} shows the simulated images for the 
low-energy configurations of substitutional-interstitial complexes. 
The simulated image of the 
lowest energy substitutional-interstitial surface complex, $s(1)i^{\rm Ga}(1)$, 
is shown in Fig.~\ref{mnga_si_combined}(a). There is a reduction in intensity of the As surface 
atom to the right of the $s(1)$ site and in the region between the
As (001) planes to the left of the $s(1)$ site. The simulated images of 
$s(1)i^{\rm Ga}(1)^{\rm *}$ [Fig.~\ref{mnga_si_combined}(c)], 
$s(2)i^{\rm As}(2)$ [Fig.~\ref{mnga_si_combined}(d)], and $s(1)i^{\rm As}(2)$ 
[Fig.~\ref{mnga_si_combined}(f)] are all similar, but the reduction 
in the simulated intensity between the As (001) planes to the
left of the $s(1)$ site is less apparent. Among these, only 
the $s(2)i^{\rm As}(2)$ complex has a ($\overline{1}$10) 
mirror-plane symmetry. All four of these simulated images are generally 
consistent with the characteristics of feature E in the experimental data of 
Fig.~\ref{expt_stm}. However, since the $s(2)i^{\rm As}(2)$ complex is much
lower in energy ($\sim$0.5 eV) than the other complexes just
described, and since there appears to be only two such features in 
Fig.~\ref{expt_stm}, we suggest that feature E is most likely a 
$s(2)i^{\rm As}(2)$ complex.
If this is correct, then the observed apparent displacement of 
the surface As atom results mainly from an electronic effect rather than a 
real atomic displacement.

The simulated image of the $s(1)i^{\rm As}(1)$ complex shown in 
Fig.~\ref{mnga_si_combined}(b) is rather different than those for the
complexes just described. This simulated image obviously has a ($\overline{1}$10) 
mirror-plane symmetry and a ``butterfly'' shape with a region of lower
intensity between the Mn atoms. Since there is no (001) mirror plane, 
the left and right wings of the butterfly are clearly different,
with the contribution from the $s(1)$ Mn more intense than 
that of the $i^{\rm As}$(1). 
Overall, these characteristics are similar to those of
feature C, which appears to involve four As surface atoms in neighboring
(001) planes. As we discuss below, there are other energetically competitive
complexes with similar simulated images, so that we postpone any conclusions regarding
feature C until all the relevant complexes have been discussed.

The simulated image of the $s(2)i^{\rm Ga}(2)$ complex shown in 
Fig.~\ref{mnga_si_combined}(e) does not have a ($\overline{1}$10) mirror 
plane symmetry. In this configuration there is
an oblate region of intensity along the As (001) plane and a reduction
in the surface As atom intensity nearest the interstitial site. On the
basis of the available STM data, we cannot rule out the possibility that
some of those features labeled B in Fig.~\ref{expt_stm} result from
$s(2)i^{\rm Ga}(2)$ complexes.

\subsubsection{Interstitial-substitutional complexes}
\begin{figure}
\resizebox{8cm}{!}{\includegraphics{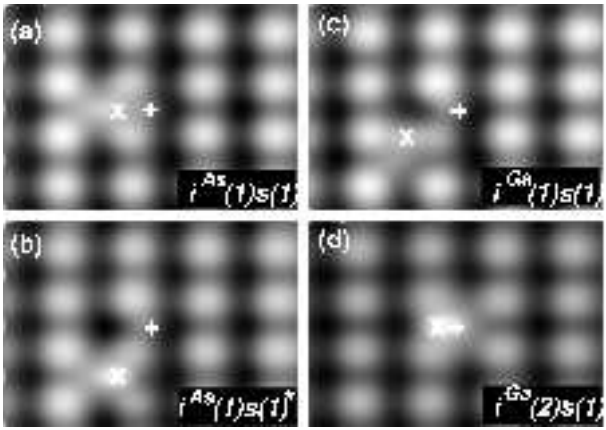}}%
\caption{\label{mnga_is_combined} Simulated STM images for interstitial-substitutional 
complexes.
The locations of substitutional and interstitial
Mn are denote with $+$ and $\times$, respectively. The crystalline orientation is 
the same as in Fig.~\ref{struk_new}.
} 
\end{figure}

Figure~\ref{mnga_is_combined} shows the simulated images for the 
low-energy configurations of interstitial-substitutional complexes. The simulated images of the
$i^{\rm As}(1)s(1)$, $i^{\rm As}(1)s(1)^{\rm *}$, and $i^{\rm Ga}(2)s(1)$
complexes are all very similar in shape, with an increase in intensity
on the four nearest surface As atoms and between the As (001) planes
to the left of the $s(1)$ site for the first two and to the right of the
$s(1)$ site for the latter. Obviously, the $i^{\rm As}(1)s(1)$ and $i^{\rm Ga}(2)s(1)$
complexes have a ($\overline{1}$10) mirror-plane symmetry, whereas the
$i^{\rm As}(1)s(1)^{\rm *}$ appears to approximately have this symmetry.
The characteristics of these images are similar to that of the 
$s(1)i^{\rm As}(1)$ complex in Fig.~\ref{mnga_si_combined}(b), but with 
more intensity between the relevant As (001) planes as a result of the
interstitial location.

The simulated image of the $i^{\rm Ga}(1)s(1)$ complex shown in 
Fig.~\ref{mnga_is_combined}(c) has no apparent symmetry. There is an
increase in intensity between the two surface As atoms nearest the
$i^{\rm Ga}(1)$ site, whereas the $s(1)$ Mn is only very weakly
visible. Features such as this are not apparent in the experimental
data, although XSTM scans of lower doped samples may help clarify
whether they are present.

From our STM data and the theoretical simulations, we cannot unambiguously
assign feature C to one particular complex among the plausible
alternatives: $s(1)i^{\rm As}(1)$, $i^{\rm As}(1)s(1)$,
$i^{\rm As}(1)s(1)^{\rm *}$, $i^{\rm Ga}(2)s(1)$, 
$s(1)s(1)$, $s(3)s(3)$ and $s(1)s(3)$. 
Considering the large number of such features in Fig.~\ref{expt_stm} 
and the comparable energies of the different complexes, it is likely
that more than one of these dopants are present in the scan area of 
Fig.~\ref{expt_stm}. Measurements on samples with much smaller Mn 
concentration are ongoing and should be able to address these issues
in more detail.

\section{Summary}\label{Results}
In summary, we have combined first-principles calculations and high-resolution
XSTM measurements to characterize Mn-doped GaAs.
XSTM on a (110) cleavage plane reveals five reproducible features 
not found in bulk GaAs. Total-energy calculations were used to screen
the most likely configuration of Mn dopants near the GaAs (110) surface.
The low-energy configurations were then further 
scrutinized by comparison of simulated and measured STM images. These
comparisons reveal that there are predominantly two types of Mn-related
dopants in the sample: (1) those involving only Mn substitutional(s), including
both isolated and pairs of substitutionals, and (2) complexes composed of substitutional 
and interstitial Mn.

\bibliographystyle{/home/main1/hellberg/bib/prsty}

\end{document}